\documentclass[reprint, 10pt,
aps, 
amsmath,amssymb,
superscriptaddress,
floatfix
]{revtex4-2}

\usepackage{circuitikz}
\usepackage{siunitx}
\usepackage{hyperref}
\usepackage{multirow}
\usepackage{comment}
\usepackage{braket}
\usepackage{soul}
\usepackage{amsmath}
\usepackage{dsfont}
\usepackage{tikz}
\usepackage{quantikz}
\usepackage{venndiagram}
\usepackage{cancel}
\usepackage{placeins}
\usepackage{float}
\usetikzlibrary{positioning, shapes.geometric}
\usepackage{amsthm}
\usepackage{needspace}

\begin{document}

\title{The Pangaea Architecture: Fault-Tolerant Heterogeneous Topological Codes via a Quantum Bus}

\author{Sheir Yarkoni}
\thanks{These authors contributed equally to this work.}
\affiliation{Qarakal Quantum, Derech Menachem Begin 144, Tel~Aviv-Yafo 6492102, Israel}

\author{Chen Scheim}
\thanks{These authors contributed equally to this work.}
\affiliation{Qarakal Quantum, Derech Menachem Begin 144, Tel~Aviv-Yafo 6492102, Israel}

\author{Daniel Hakshuri}
\affiliation{Qarakal Quantum, Derech Menachem Begin 144, Tel~Aviv-Yafo 6492102, Israel}

\author{Nadav Katz}
\affiliation{Qarakal Quantum, Derech Menachem Begin 144, Tel~Aviv-Yafo 6492102, Israel}
\affiliation{Racah Institute of Physics, The Hebrew University of Jerusalem, Jerusalem 9190401, Israel}

\begin{abstract}
We introduce Pangaea, a fault-tolerant quantum architecture that uses a quantum bus to mediate logical operations between remote patches of two-dimensional topological codes. The bus is an auxiliary gauge-code strip whose measurements reconstruct joint logical operators while preserving nearest-neighbor physical connectivity. Enabling native heterogeneous topological codes and multi-qubit Pauli operations, the quantum bus can be interpreted as a three-dimensional generalization of lattice surgery. We require only \(O(dN_L)\) physical qubits to implement multi-qubit interactions for \(N_L\) distance-\(d\) logical qubits, compared to $O(d^2N_L)$ of traditional two-dimensional architectures. At the 50-logical-qubit scale, Pangaea uses up to \(10\times\) fewer physical qubits than planar surface-code architectures at matched logical error rates. We verify fault-tolerance of long-range measurement-based CNOT primitives for both surface--surface and surface--color joint parity measurements using pseudo-threshold simulations. We use this protocol to construct a native heterogeneous 15-to-1 magic-state distillation module using the quantum bus. These results establish Pangaea as a scalable architecture for three-dimensional fault-tolerant quantum computing that resolves the routing bottleneck of planar lattice surgery.

\end{abstract}

\maketitle
\enlargethispage{3\baselineskip}
\vspace{-0.6\baselineskip}

\section{Introduction}

Quantum error correction (QEC) enables fault-tolerant quantum computation by encoding quantum information across many qubits, such that errors can be inferred and corrected without measuring the logical state. A logical qubit is typically encoded on a physical set of data and ancilla qubits forming a \textit{code patch}. Different QEC codes, such as surface-codes~\cite{Fowler2012SurfaceCodes, Dennis2002, Marques2022}, color-codes~\cite{Bombin2006, LitinskiKesselring2017, ThomsenKesselring2022}, and quantum low-density parity-check (qLDPC) codes~\cite{Yoder2025, Gu2026NearestNeighbour}, define these patches through distinct stabilizer structures~\cite{Gottesman1997Stabilizer}, leading to different overheads, thresholds, connectivity requirements, and sets of native logical operations~\cite{Dennis2002,Litinski2019,Bombin2006,Yoder2025, GoogleQuantumAI2025BelowThreshold}.
Superconducting qubits are among the leading platforms for realizing these codes, with recent experiments having demonstrated logical state preparation, error suppression, teleportation-based operations, measurement-induced two-qubit gates, and near-term lattice-surgery entanglement protocols~\cite{Besedin2026,Lacroix2025,Bodeker2026}. 

These results establish the feasibility of QEC building blocks, but scalable fault-tolerant processors require logical operations among many encoded qubits, often across non-neighboring or functionally different modules---a challenge made more acute by the connectivity limitations of superconducting hardware~\cite{DiVincenzo2009}. Physical-qubit connectivity therefore represents a trade-off against the efficiency of the logical operations it can support, with recent advances expanding the capabilities of connections in superconducting hardware~\cite{Renger2026AllToAll}.

Existing approaches solve this connectivity problem in complementary but limited ways. Some QEC codes admit transversal multi-qubit gate operations, but come at the cost of increasing the physical qubit connectivity requirements~\cite{BravyiCross2024}. Defect-based topological schemes entangle many logical qubits by braiding or code deformation, effectively moving information through the logical qubit network, but require more complex scheduling and decoding schemes~\cite{RaussendorfHarrington2007, Fowler2012SurfaceCodes, Fujiu2026DensePacking, Vuillot2019GaugeFixing}. Lattice surgery avoids a direct physical two-qubit gate by merging and splitting adjacent logical qubits' joint operators, making it one of the most practical QEC primitives~\cite{Horsman2012,Litinski2019,Bodeker2026}. However, non-adjacent patches must be routed together, incurring a high physical qubit overhead~\cite{Litinski2019}, and surgery protocols are typically designed for homogeneous code architectures~\cite{Litinski2019, Choe2024, BravyiCross2024, CohenTan2022, Fowler2010SurfaceCodeComm, Ha2025LatticeSurgery, Hamada2026EfficientRouting}. To address this, basic heterogeneous interfaces have been developed such as surface--color-code switching, merged surface--color decoding~\cite{Nautrup2017,BeverlandKubica2021,Shutty2022,Herzog2025}, and teleportation-based architectures~\cite{Choe2024}. High-rate qLDPC proposals also use heterogeneous codes as adapters for logical gates~\cite{Yoder2025,Webster2026Pinnacle}, although such architectures also require significant qubit connectivity overhead.

What is still missing is a native fault-tolerant interaction primitive that simultaneously supports long-range connectivity, heterogeneous code patches, and a common scalable layout, all present in a single architecture. We address this gap by proposing a heterogeneous, three-dimensional fault-tolerant quantum architecture we call Pangaea, where long-range logical connectivity is achieved via a \textit{quantum bus}; a three-dimensional extension of lattice surgery. The bus is an auxiliary gauge-code patch connecting many logical qubits through a contiguous qubit patch~\cite{Brennen2003QuantumComputerArchitecture}. Motivated by superconducting qubits, the bus requires nearest-neighbor and up to degree-6 physical qubit connectivity, compatible with existing routing and fabrication requirements~\cite{Wang2026LowOverhead, BravyiCross2024, Mathews2026Routing}.

The quantum bus serves two purposes: firstly as a configurable interface between logical qubits, and secondly as a native interface between heterogeneous code patches. The bus's gauge degrees of freedom are fixed to match the boundary stabilizers of the logical qubits' code patches, and products of local gauge-check outcomes give the desired joint logical observable, similar to traditional measurement-based primitives~\cite{Horsman2012,Lacroix2025,Shutty2022,Herzog2025,Riste2013,GoogleQuantumAI2023}. The bus stores no logical information and can couple patches that differ in code family, geometry, or boundary type, resulting in a measurement-based surgery primitive between heterogeneous topological codes. Our quantum bus connects multiple logical qubits, enabling both all-to-all logical connectivity and multi-qubit Pauli operations. By defining the bus as an extension of lattice surgery, the bus serves as a logical primitive which inherently preserves fault-tolerance, as opposed to competing proposals for long-range interconnects which rely on either lossy channels~\cite{Kurpiers2018QuantumBus, Axline2018Interconnect} or noisy physical connections~\cite{Perseguers2010,Wang2026LowOverhead}.

Pangaea's use of a three-dimensional interconnect to both scale system size and mediate logical interactions distinguishes it from other three-dimensional architectures, such as qLDPC~\cite{BravyiCross2024, KovalevPryadko2013,Yoder2025}, toric code~\cite{Dennis2002}, or 3D color-code-based architectures~\cite{Butt2026Complementary3DColor}, which reduce qubit overhead by replacing two-dimensional topological codes with higher-rate codes at the cost of longer-range physical connectivity. Such proposals are promising in terms of encoding rate but require complex connectivity~\cite{BravyiCross2024, KovalevPryadko2013}, measurement scheduling~\cite{Yoder2025}, and decoding and operation~\cite{Roffe2020BPOSD}. Pangaea, however, retains nearest-neighbor connectivity throughout the entirety of the architecture, simplifying construction and connection of differing QEC code types. Our aspiration is to provide a practical architecture that achieves fault tolerance in large-scale quantum computers. The main contributions of our work are:

\needspace{5\baselineskip}
\begin{itemize}
    \item \textbf{Heterogeneous topological-code integration.}
    Pangaea's quantum bus provides a common lattice surgery-like interface between code families with different stabilizer structures. This enables logical operations to be defined across heterogeneous QEC codes rather than within a single code family. We demonstrate fault-tolerance explicitly for surface--color logical operations implemented within a single bus.
    
    \item \textbf{Reduced physical-qubit overhead for modular scaling.}
    The Pangaea quantum bus requires only \(O(dN_L)\) physical qubits to mediate fault-tolerant interactions for $N_L$ distance-\(d\) logical qubits, compared with the \(O(d^2N_L)\) ancillary overhead of conventional two-dimensional layouts. Because these operations do not require the participating code patches to be expanded, contracted, or rerouted, each logical-qubit footprint remains fixed. Additional qubits can therefore be integrated by extending the bus rather than redesigning the surrounding interaction region.

    \item \textbf{Long-range and multi-qubit logical parity measurements.}
    In Pangaea, dressed Pauli observables between remote code patches can be measured as products of local gauge checks on the bus. This construction natively extends two-qubit joint-parity measurements to multi-qubit Pauli measurements, avoiding decomposition into pairwise operations or large code deformations. We illustrate this capability with a heterogeneous implementation of the 15-to-1 magic-state distillation protocol.

\end{itemize}

\section{Quantum Bus for Logical Qubit Interactions}
\label{sec:quantum_bus_intro}

At the physical-qubit level, superconducting processors have already used measurement as an entangling resource: measurement with feedforward can generate two-qubit entanglement~\cite{Riste2013}, dynamic circuits enable gate teleportation and create long-range interactions through mid-circuit measurements~\cite{Baumer2024}, and monitored-circuit experiments have observed measurement-induced entanglement and teleportation phenomena~\cite{GoogleQuantumAI2023}. 

In quantum error correction, repeated stabilizer measurements along a boundary between two code patches can be used to infer the joint parity between two logical qubits. This is lattice surgery, which enables teleportation, entanglement, or logical gates between code patches~\cite{Horsman2012,Besedin2026,Wang2026,Bodeker2026}. Pangaea's quantum bus extends this notion, providing a standardized measurement-based protocol for many-logical-qubit operations as a basis for a complete fault-tolerant architecture.

We introduce Pangaea's quantum bus in three stages. Sec.~\ref{subsec:bus_minimal} re-formulates lattice surgery as the minimal instance of a quantum bus, allowing joint parity measurements of two code patches. Sec.~\ref{subsec:bus_operation} generalizes the bus's core primitive, showing how physically extending the bus still allows fault-tolerant joint parity measurements at long distances. Finally, Sec.~\ref{sec:mb_cnot_ls} shows how two such measurements, combined with an ancilla logical qubit, realize a full two-qubit gate: the measurement-based CNOT.

\subsection{Lattice Surgery as a Minimal Quantum Bus: The Scissors Operator}
\label{subsec:bus_minimal}
\begin{figure*}[t]
    \centering
    \includegraphics[width=0.9\textwidth]{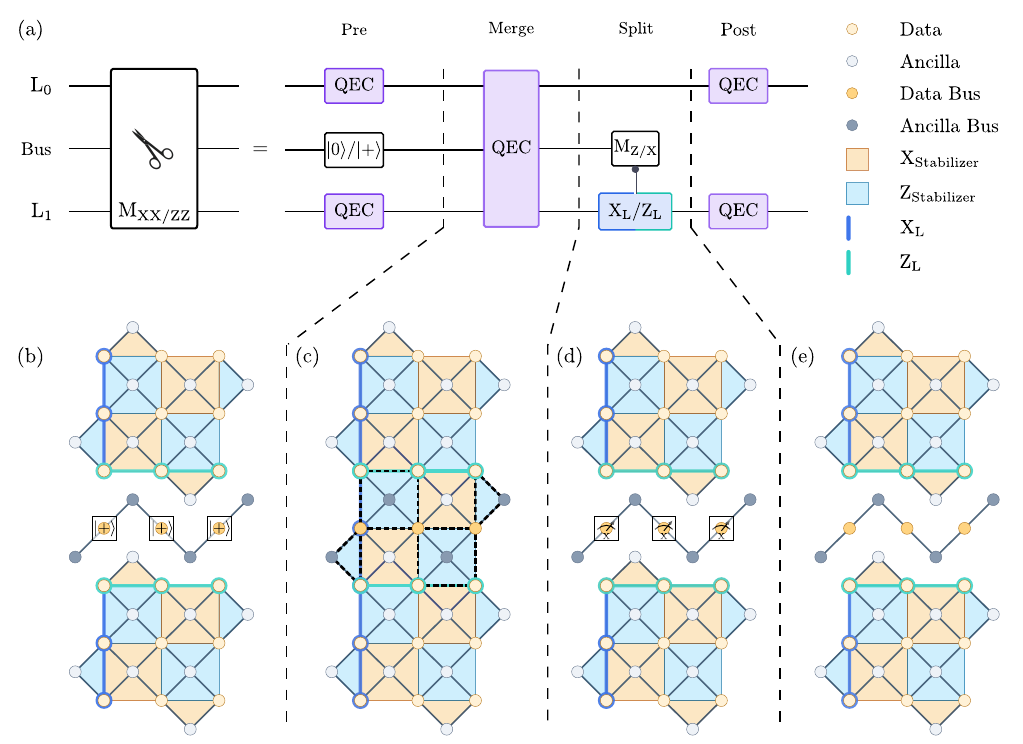}
    \caption{
    Lattice surgery between two distance-3 rotated surface-code patches via a quantum bus, shown explicitly for an $M_{ZZ}$ interaction.
    \textbf{(a)} Circuit-level diagram defining the main stages of lattice surgery, including fault-tolerant stabilizer measurements, which for future use we denote as the ``scissors'' operator, composed of: pre-op, merge, split, and post-op. \textbf{(b)-(e)} Equivalent schematic diagrams defining the four stages of the scissors operator. Data qubits participating in each logical $X_L (Z_L)$ operator are highlighted in blue (turquoise).}
    \label{fig:lattice_surgery_scheme}
\end{figure*}

We briefly review the
protocol for lattice surgery (shown in Fig.~\ref{fig:lattice_surgery_scheme}) and extend it to a fundamental primitive in the Pangaea architecture. We call the
entire end-to-end fault-tolerant surgery protocol the ``scissors'' operation. Given two logical code patches, the standard auxiliary qubit strip in lattice surgery is used to mediate the interaction. We consider this single chain strip as the minimal instance of our quantum bus. For the technical details of lattice surgery, we refer the reader to Refs.~\cite{Horsman2012, Herzog2025, Litinski2019, Nautrup2017,
Besedin2026}.

The scissors operator proceeds through four stages: pre-op, merge, split,
and post-op. This process is shown at the circuit level in Fig.~\ref{fig:lattice_surgery_scheme}(a)
and schematically in Figs.~\ref{fig:lattice_surgery_scheme}(b)--(e) for an $M_{ZZ}$ measurement. Before and after the operation---Figs.~\ref{fig:lattice_surgery_scheme}(b) and (e)---the two logical qubits are stabilized independently, each by its own check cycles. During the merge, the bus and the two patches form a single contiguous patch whose bus-spanning stabilizers reproduce the joint parity measurement. The dashed \(Z\)-type stabilizers in Fig.~\ref{fig:lattice_surgery_scheme}(c) form the generator set \(\mathcal{G}_{B}^{(Z)}\). Their measurement outcomes determine the joint parity of the two logical \(Z\) operators, projecting the
two-qubit logical state into the corresponding joint-parity subspace. Formally, for two logical operators represented by participating boundary $\overline{Z}_{\partial}^{(L_0)}$ and $\overline{Z}_{\partial}^{(L_1)}$, the merge stabilizer generators satisfy

\begin{equation}
    \prod_{g \in \mathcal{G}_{B}^{(Z)}} g
    =
    \overline{Z}_{\partial}^{(L_0)} \overline{Z}_{\partial}^{(L_1)} S^{(Z)}_B,
\end{equation}
where $S_B^{(Z)}$ is the bus stabilizer product. Together, these form the joint logical operator $\overline{Z}^{(L_0)}\overline{Z}^{(L_1)}$. The joint $M_{ZZ}$ can be measured as
\begin{equation}
    m_{ZZ}
    = s_B\prod_{g\in\mathcal{G}_{B}^{(Z)}} m_g ,
    \label{eq:minimal_bus_parity}
\end{equation}
where \(m_g\in\{\pm1\}\) is the measured outcome of the merge stabilizer \(g\) and $s_B\in\{\pm1\}$ is the known eigenvalue of \(S_B^{(Z)}\). The split measures out the bus data qubits, and the resulting outcomes fix a conditional Pauli-frame correction, shown in Fig.~\ref{fig:lattice_surgery_scheme}. As in standard lattice surgery, this protocol holds for both $M_{ZZ}$ and $M_{XX}$ with appropriate changes. 

\subsection{From Lattice Surgery to a Quantum Bus}
\label{subsec:bus_operation}

We now show how to extend lattice surgery to an arbitrary-distance bus-mediated logical operation. Consider the lattice surgery mediating strip shown between the two code patches in Fig.~\ref{fig:lattice_surgery_scheme}. In two-dimensional architectures, this represents a \textit{single} logical interaction between two directly adjacent code patches. We take this mediating strip and stretch it vertically, with the intention of connecting \textit{many} logical qubits. Concretely, we replicate many rows of the mediating patch, stack them vertically, and connect them as a contiguous mediating strip to which logical qubit code patches will connect. This is our definition of the quantum bus.

We define the bus length, \(\ell\), as the number of said data qubit rows in the bus. Thus, the bus can connect a total number of $N_L = \ell+1$ logical qubits. Note that during a multi-qubit operation the bus provides distance-$d$ protection of the joint parity measurement, and a simultaneous protection of up to $2d+\ell$ in the complementary basis. For $\ell=1$ we recover lattice surgery with $N_L=2$ logical qubits. 

A key difference between Pangaea and two-dimensional architectures is the number of physical qubits required to mediate long-range interactions. Traditional lattice surgery-based implementations use $O(d^2N_L)$ qubits to construct intermediary patches that connect the logical qubits' operators~\cite{Litinski2019, Choe2024}. By defining the quantum bus orthogonally to the code patches and vertically stacking the code patches, we achieve that (i) each logical qubit's operator aligns directly with the bus, allowing lattice surgery between immediately adjacent qubits, and (ii) the ancilla qubits of each code patch touching the quantum bus are reused for the gauge generators of the bus. This significantly reduces the physical qubits required to implement fault-tolerant logical operations. For distance-$d$ logical qubits, logical fault-tolerance can be maintained with a bus that is $d$ qubits wide, and $\ell$ data rows long. Thus, the total number of qubits in the quantum bus scales as $O(dN_L)$---a factor $d$ improvement over known two-dimensional architectures. A full accounting derivation is given in 
Appendix~\ref{app:physical_qubit_count}.

Fig.~\ref{fig:lattice_surgery_scheme_3d} shows two orthogonal quantum buses arranged in a three-dimensional tower, with code patches inserted into dedicated bus slots. Yellow and gray nodes are additional rows of bus data and ancilla qubits respectively, seen in Fig.~\ref{fig:lattice_surgery_scheme_3d}(a). The bus is defined so that code patches' rough boundaries ancillas act as the complementary check qubits of the bus. This allows us to ``insert'' the rough boundaries of code patches into the highlighted \textit{bus slots}, shown as dashed line nodes and boxes. The bus reuses the patches' ancillas to perform stabilizer measurements with connectivity shown in Fig.~\ref{fig:lattice_surgery_scheme_3d}(b). Code patches expose their logical \(Z_L\) and \(X_L\) boundaries towards the bus, effectively integrating directly into the bus. Code patch ancilla qubits connect to the rows of bus data qubits from above and below, whereas the data qubits connect to the bus ancilla qubits in the same row. This connectivity is repeated for every additional code patch inserted into the bus, creating a contiguous, common interface between the quantum bus and all code patches connected to it. 

\begin{figure}[htbp]
    \centering
    \includegraphics[width=\columnwidth]{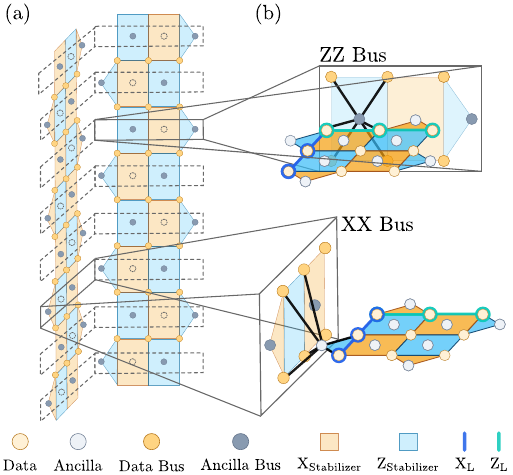}
    \caption{
    Three-dimensional layout of two orthogonal quantum buses and their physical connections to two non-adjacent logical qubits. 
    \textbf{(a)} Two orthogonal bus strips form a three-dimensional tower to enable long-distance interactions. Dashed lines highlight the respective ``bus slots'', where each slot accepts a single code patch. Yellow nodes denote bus data qubits, gray nodes are ancilla qubits, and dashed nodes are ``missing'' qubits where rough boundaries of the code patch are inserted into. The $ZZ$-bus ($XX$-bus) is aligned with each surface-code patch's logical $Z_L$ ($X_L$) operator for native $M_{ZZ}$ ($M_{XX}$) parity measurements. 
    \textbf{(b)} Zoomed-in diagram emphasizing the code patch-to-bus connectivity which allows joint stabilizer measurements---logical $Z_L$ ($X_L$) is shown on top (bottom).
    }
    \label{fig:lattice_surgery_scheme_3d}
\end{figure}

Long-range \(M_{ZZ}/M_{XX}\) measurements through the bus are obtained by extending the mediating region of the lattice surgery merge across multiple bus rows. For the same $M_{ZZ}$ example as before (Fig.~\ref{fig:lattice_surgery_scheme}), the set of \(Z\)-type merge-check generators is now associated with an additional index, the bus row \(k\). The bus generator set is therefore denoted as \(\mathcal{G}_{B,k}^{(Z)}\), with operator
\begin{equation}
    g_k
    \equiv
    \prod_{g\in\mathcal{G}_{B,k}^{(Z)}} g .
\end{equation}
The complete bus merge-generator set becomes
\begin{equation}
    \mathcal{G}_{B}^{(Z)}
    =
    \bigcup_{k=1}^{\ell}\mathcal{G}_{B,k}^{(Z)},
\end{equation}
where each additional bus row contributes a distinct factor \(g_k\) to the measured operator. Multiplying the row operators across the full bus gives
\begin{equation}
    \prod_{k=1}^{\ell} g_k
    =
    \prod_{k=1}^{\ell}
    \prod_{g\in\mathcal{G}_{B,k}^{(Z)}} g
    =
    \overline{Z}_{\partial}^{(L_0)}
    \overline{Z}_{\partial}^{(L_1)}
    S^{(Z)}_B ,
    \label{eq:extended_bus_operator}
\end{equation}
with \(S^{(Z)}_B\) again being the product of the appropriate stabilizers within the mediating bus region, consistent with our quantum bus definition of lattice surgery. In long-range interactions, the intermediate Pauli factors cancel between adjacent rows, leaving only the two boundary logical operators, dressed by \(S_B\). The \(M_{XX}\) measurement follows identically upon exchanging \(Z\leftrightarrow X\). 

Note that we still retain the ability to perform long-distance, many-logical-qubit Pauli operations through the bus. This follows directly from the definition of the gauge generators, where we can expose any number of Pauli operators; this yields native many-logical-qubit joint parity measurements under the same construction introduced here. The only criterion is that the boundaries of the logical qubits participating in the multi-logical-qubit operations are connected to the bus, where the joint operator can be expressed as in Eq.~\eqref{eq:extended_bus_operator}, and similarly for $X$-type interactions. However, in this specific case, the stabilizer along the boundary between the code patch and the bus introduces a check that does not commute with its neighbors. We resolve this by decomposing the check into two consecutive lower-weight checks that overlap on even numbers of data qubits, and both commute with the neighboring checks, thus resolving the anti-commutation. This allows us to retain one of the main strengths of two-dimensional architectures, where such operations are native. We show an illustration of a native three-qubit joint $M_{XXX}$ operation, its respective commutation correction, and threshold results in Appendix~\ref{app:multi_terminal_bus}.

\subsection{Measurement-Based CNOT: A Logical Two-Qubit Gate}
\label{sec:mb_cnot_ls}

\begin{figure}[h!]
    \centering
    \includegraphics[width=0.75\columnwidth]{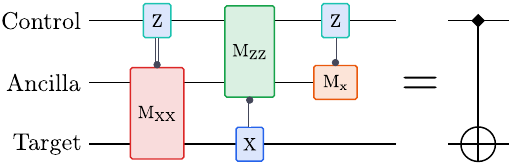}
    \caption{
    Circuit-level description of a measurement-based CNOT using joint parity measurements. An ancilla mediates the gate between the control and target qubits. Red denotes the \(M_{XX}\) measurement between ancilla and target, green denotes the \(M_{ZZ}\) measurement between control and ancilla, orange denotes the final ancilla \(M_X\) measurement, and blue denotes the relevant conditional Pauli-frame updates.
    }
    \label{fig:mb_cnot_simple}
\end{figure}

Logical measurement-based CNOTs require two joint parity measurements: $M_{ZZ}$ and $M_{XX}$. Having established how the bus mediates a single joint parity measurement at arbitrary bus length, we now show how to realize a full two-qubit measurement-based CNOT gate in Pangaea; as in standard lattice surgery, we require a control qubit, target qubit, and ancilla qubit~\cite{Horsman2012,Litinski2019,Bodeker2026}. The protocol is summarized in Fig.~\ref{fig:mb_cnot_simple}, whose color coding for these four operations is used consistently throughout the rest of this paper.

\begin{figure}[htbp]
    \centering
    \includegraphics[width=\columnwidth]{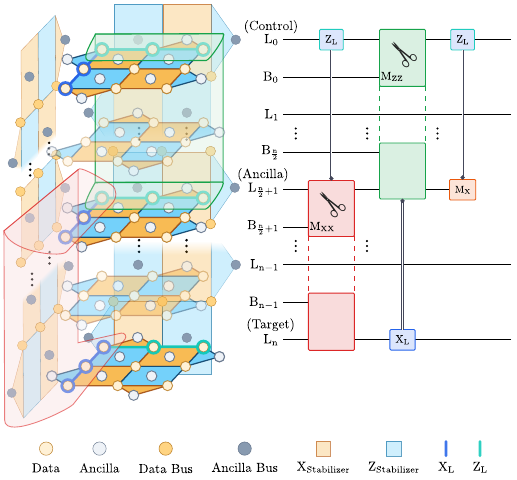}
    \caption{
    Full measurement-based CNOT between non-adjacent logical qubits using the quantum bus, realized as a sequence of joint parity measurements ($M_{XX}$, $M_{ZZ}$) followed by an ancilla readout, built from operational primitives defined in Fig.~\ref{fig:lattice_surgery_scheme}. Each colored wrapping strip on the left side illustrates the scissors operator on the relevant logical qubit patches. 
    }
    \label{fig:lattice_surgery_scheme_cnot}
\end{figure}

Fig.~\ref{fig:lattice_surgery_scheme_cnot} shows a schematic of a bus-mediated CNOT. We implement observables \(M_{XX} = \overline{X}^{\rm (Ancilla)}\overline{X}^{\rm (Target)}\) and \(M_{ZZ} = \overline{Z}^{\rm (Ancilla)}\overline{Z}^{\rm (Control)}\), defined through the bus. The logical ancilla is prepared in the logical \(\ket{0_L}\) state, so that after these measurements it can be read out in the \(X\) basis,
denoted \(M_X\), and the three measurements together determine Pauli-frame updates on the control and target.

This further solidifies the quantum bus as a direct three-dimensional extension to lattice surgery: Fig.~\ref{fig:lattice_surgery_scheme_cnot} is equivalent to
Fig.~\ref{fig:mb_cnot_simple}, with the scissors operator as defined in
Fig.~\ref{fig:lattice_surgery_scheme}(a). The bus improves upon this with the added ability to perform the same measurement-based CNOT between two logical qubits that are not \emph{physically} adjacent by routing the joint parity measurement through the bus---rather than a single merged boundary---using $O(dN_L)$ physical qubits.

\label{subsec:color_codes_integration}
\begin{figure*}[t]
    \centering
    \includegraphics[width=\textwidth]{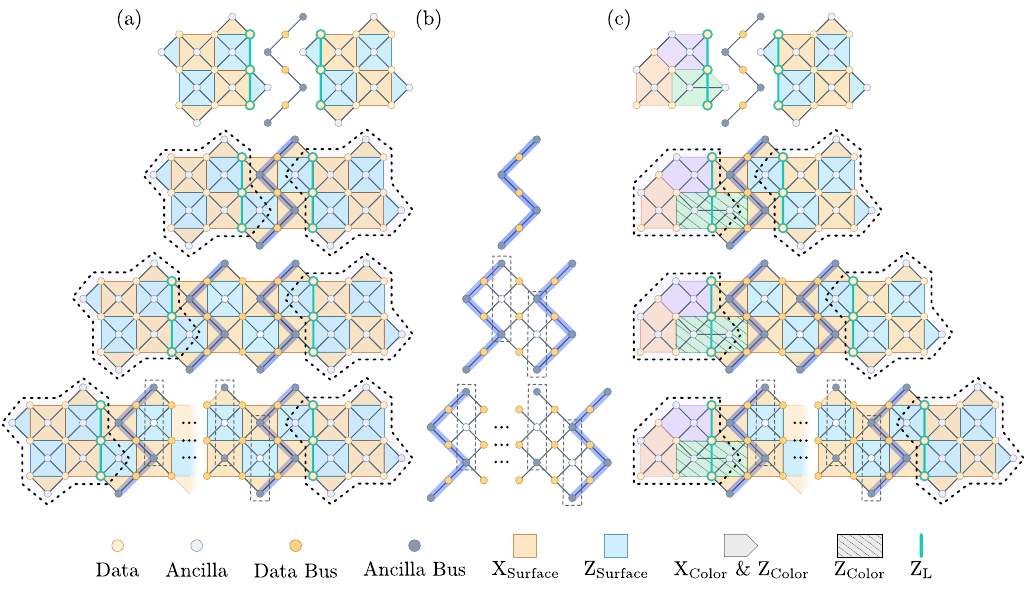}
    \caption{
    Bird's-eye view of a single bus strip connecting surface-to-surface (left) and surface-to-color (right) patches.
    \textbf{(a)} Top to bottom: a bus connecting two adjacent patches (as in Fig.~\ref{fig:lattice_surgery_scheme}) is extended to accommodate additional slots; bottom row shows an arbitrary-length extended bus connecting distant code patches. \textbf{(b)} Increasing the length of the bus naturally creates additional slots for logical qubit insertion (shaded blue lines mark the endpoints). Bottom row shows the modularity and composability of the bus structure; additional logical qubit slots are highlighted with dashed boxes. \textbf{(c)} Surface-to-color interface with growing bus length as in (a). The bus structure natively supports both surface and color-code patches with modular interchangeability to enable long-range entanglement.
    }
    \label{fig:bus_growth}
\end{figure*}

\section{Fault-Tolerance of Heterogeneous Code Mixing}
\label{sec:bus_growth}

Until now, we have used the surface-code as our code of choice to explain the bus's functionality. We now show how the Pangaea architecture natively integrates color-code qubits, standardizing our quantum bus as an interface between different code types. Integrating heterogeneous codes is generally nontrivial because surface-codes and color-codes have different stabilizer structures, boundary geometries, and native logical capabilities~\cite{Nautrup2017,Hirano2025Efficient}. In Pangaea, we require only that the boundary between the bus and the code of choice be prepared in such a way that the appropriate logical Pauli operator is exposed as a rough surface to integrate into a slot in the bus. Establishing fault-tolerance of the quantum bus requires that (i) errors within the code do not go undetected, and (ii) no additional logical degrees of freedom are created by the bus. For condition (i) we perform numerical simulations of noisy circuits to directly extract pseudo-threshold values at various bus lengths. For condition (ii) we explicitly count the degrees of freedom and enumerate the stabilizers in all stages of the scissor operator in Appendix~\ref{app:gauge-fixing-bus}, and show that no residual logical degrees of freedom exist.

We show fault tolerance of using the scissors operator for heterogeneous codes: Sec.~\ref{subsec:color_code_nkd} shows how color-code qubits are integrated into the quantum bus, and formalizes the merged system in $[[n, k, d]]$ notation. Sec.~\ref{subsec:threshold} shows results of numerical simulations for both homogeneous and heterogeneous use of the quantum bus, explicitly calculating pseudo-thresholds and extracting logical error rates of the bus. We then compare Pangaea to two-dimensional surface-code architectures and estimate the resources required to achieve fault-tolerance in Sec.~\ref{subsec:qubit_budget_analysis}.

\subsection{Integrating Color-Codes into the Quantum Bus}
\label{subsec:color_code_nkd}
In general, any two-dimensional code patch with distance $d \leq d_{\mathrm{bus}}$ can be supported natively by the bus~\cite{Hirano2025Efficient}. For simplicity, we show how code patches matching the width \(d_{\mathrm{bus}}\) are inserted. We show logical operations via a flattened, elongated bus for both surface-to-surface patches in Fig.~\ref{fig:bus_growth}(a) and surface-to-color patches in Fig.~\ref{fig:bus_growth}(c). Fig.~\ref{fig:bus_growth}(b) highlights a fundamental property of the bus; repeating the structure of the bus (extending the length to $\ell>1$) naturally exposes additional slots, allowing insertion of additional code patches. The bus mediates between heterogeneous codes by applying the four stages of the scissors operator to subsystem lattice surgery~\cite{Nautrup2017}. We explicitly show how to connect color-code patches to the quantum bus in Fig.~\ref{fig:bus_growth}(c) by defining the boundary of the patch so it matches the structure of the rough boundary of a surface-code patch. We use the superdense implementation of color-code~\cite{Lacroix2025} so that we can create this rough boundary along the logical operator of each color-code patch. This allows each patch to connect to the bus by extending weight-4 stabilizers to weight-6 with the same structure as the weight-2 stabilizers in surface-code extend to weight-4. Although this does not conform to a square grid, this repeated structure obeys degree-4 connectivity within each logical qubit and repeats throughout the color-code patches as code distance grows.

\begin{figure*}[t]
    \centering
    \includegraphics[width=\textwidth]{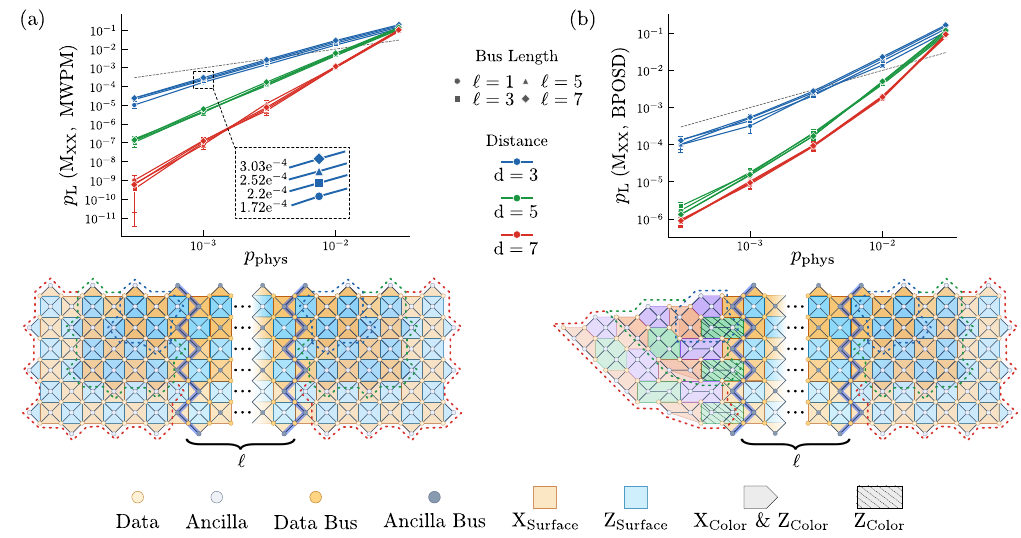}
    \caption{
    Pseudo-threshold simulations for bus-mediated \(\mathrm{M_{XX}}\) parity measurements. Dashed lines represent the $p_L=p_{phys}$ crossover point in each plot. \textbf{(a)} Logical error rate \(p_\mathrm{L}\) versus physical error rate \(\mathrm{p_{phys}}\) for distances \(d=3,5,7\), for homogeneous surface--surface bus operations decoded with minimum-weight perfect matching (MWPM). \textbf{(b)} Simulated error rates of heterogeneous surface--color bus operations decoded with a belief-propagation decoder (BPOSD). For both, the pre-op--merge--post-op circuit construction uses a round-level depolarizing plus measurement-error model as described in the text. Error bars are normal-approximation \(95\%\) confidence intervals, and the quoted pseudo-thresholds \(p^*\) are finite-distance crossing estimates obtained by linear interpolation between the sampled points where the sign of \(p_\mathrm{L}-p_\mathrm{phys}\) changes.}
    \label{fig:thresholds}
\end{figure*}

To formalize the merge operation of multiple code patches of different code types, we use the $[[n,k,d]]$~\cite{Gottesman1997Stabilizer} formalism to track the logical degrees of freedom throughout our code structure. Here, $n$ is the total number of data qubits, $k$ is the number of logical qubits and $d$ is the code distance. In standard subsystem lattice surgery, to ensure commutation of the check cycles of color-code patches along the merge boundary, one check must be dropped~\cite{Nautrup2017}. We show this in Fig.~\ref{fig:bus_growth}(c), where a single green color-code tile is hash-filled, representing only a single-basis check. Dropping this check introduces a gauge degree of freedom, reinterpreted in this context as a subsystem code, $[[n,k,r,d]]$, where $r$ is the number of gauge degrees of freedom. We derive the $[[n, k, r, d]]$ expressions for both split and merge stages of bus-mediated lattice surgery for the three possible combinations: the two homogeneous cases, all-surface and all-color, and the heterogeneous surface-color case (see Appendix~\ref{app:gauge-fixing-bus} for details). 

For $N_L$ logical qubits, with $\mu$ number of distance-$d$ color-code patches and $N_L-\mu$ surface-code patches, we denote $n_{\rm split}$ as the sum of data qubits from the independent code patches, and $n_{\rm merge}$ as the total number of data qubits during the merge. Accounting for the bus data qubits, we obtain a total of $n_{\mathrm{merge}}=n_{\mathrm{split}}+d\ell$. We provide a thorough derivation in Appendix~\ref{app:gauge-fixing-bus}. The merged system is a code with parameters
\begin{equation}
[[n_{\mathrm{merge}},N_L-1,\mu(d-1)/2,d]].
\end{equation}

Note that for the two homogeneous cases: $\mu=0$ recovers $[[n_{\mathrm{merge}},\,N_L-1,\,d]]$ (standard lattice surgery), and $\mu=N_L$ results in $[[n_{\mathrm{merge}},\,N_L-1,\,N_L(d-1)/2,\,d]]$ (subsystem lattice surgery). This formalizes the stabilizer notation for surface-code and color-code configurations of the quantum bus.

\subsection{Threshold Analysis}
\label{subsec:threshold}
We now show explicitly through numerical simulations that fault-tolerance holds for the bus. This analysis aims to verify that varying code distances and bus lengths do not significantly deteriorate the ability to detect and correct errors using the bus. We specifically test our surface-to-surface and surface-to-color layouts, with threshold simulations below verifying representative instances under a realistic noise model. We use the \textit{pseudo-threshold} $p^*$---the critical physical error rate at which the logical and physical error rates are equal---to empirically evaluate fault-tolerance. 

We present results of numerical simulations for the homogeneous surface-to-surface case in Fig.~\ref{fig:thresholds}(a) and heterogeneous surface-to-color interactions in Fig.~\ref{fig:thresholds}(b). A higher pseudo-threshold indicates the code can tolerate noisier hardware. Below the pseudo-threshold, the logical error rate falls below the physical error rate at that distance; above it, logical errors outpace the physical noise and the code no longer offers protection. For homogeneous surface-to-surface buses, pseudo-thresholds are $p^* = 0.43\%$ ($d$=3), $1.13\%$ ($d$=5), and $1.27\%$ ($d$=7); for heterogeneous surface-to-color buses, the corresponding values are $0.36\%$ ($d$=3), $1.13\%$ ($d$=5), and $1.25\%$ ($d$=7). In Fig.~\ref{fig:thresholds}, we show the logical error rate $p_\mathrm{L}$ as a function of physical error rate $p_{\rm phys}$. 

The monotonic increase in pseudo-threshold with distance corresponds to the monotonic decrease in logical error rate---both show that our quantum bus indeed preserves fault-tolerance as length increases, extending the concept of preservation in standard lattice-surgery implementations~\cite{Bodeker2026}. Concretely,   Fig.~\ref{fig:thresholds} shows that increasing $\ell$ from 1 to 7 does not pose a threat to fault tolerance. As expected, we see that the logical error rates per $p_{\rm phys}$ are ordered by bus length where longer $\ell$ induces more errors. By definition, the bus retains the ability to correct up to $\lfloor(d-1)/2\rfloor$ errors in the merged patch.   

Each data point is obtained from an explicit detector-error-model circuit for a bus-mediated \(M_{XX}\) primitive, with the assumption that the result also holds for $M_{ZZ}$ under a simple exchange of observables. The circuit is divided into three phases corresponding to the phases introduced in Fig.~\ref{fig:lattice_surgery_scheme}: the \textit{pre-op} phase in which the two qubits are measured independently, the \textit{merge} phase in which the bus qubits actively participate in calculating the joint \(M_{XX}\) observable, and the \textit{post-op} phase in which the bus is removed and the original independent qubit stabilizers are restored. For the distance-\(d\) threshold curves, we use \(r_{\rm pre-op}=r_{\rm merge}=r_{\rm post-op}=d\) syndrome extraction rounds. We simulate distances $3$, $5$ and $7$ and bus lengths $1$, $3$, $5$ and $7$ for both protocols.

We use the Stim~\cite{Gidney2021Stim} software package to perform our simulations, with physical error rate \(p_{\rm phys}\) applied uniformly throughout the circuit model for both homogeneous and heterogeneous interactions. The same verification criteria are established for both cases---no undetected single-fault logical error, and a unique detection-event syndrome per single-fault equivalence class (see Appendix~\ref{app:detector_model} for details). After initialization and before every syndrome round, all physical qubits undergo independent single-qubit depolarizing noise with probability \(p_{\rm phys}\). Each multi-qubit Pauli measurement after initialization is measured with an independent outcome-flip probability \(p_{\rm phys}\). In Fig.~\ref{fig:thresholds}, the bus is initialized in the product \(\ket{+}\) state. Reset operations and the final destructive \(X\)-basis data readout are treated as ideal. Surface-to-surface circuits in Fig.~\ref{fig:thresholds}(a) were decoded using minimum-weight perfect matching (MWPM), the standard for surface-code error detection~\cite{Fowler2015MWPM}. Fig.~\ref{fig:thresholds}(b) shows threshold calculations for surface-to-color circuits using a belief-propagation plus ordered-statistics decoder (BPOSD)~\cite{Roffe2020BPOSD, Koutsioumpas2025VibeLSD}. More details on detector model and decoder hyperparameter choices is given in Appendix~\ref{app:detector_model}.

\subsection{Qubit Budget Analysis}
\label{subsec:qubit_budget_analysis}

We now evaluate the physical-qubit cost of achieving fault-tolerance with the Pangaea architecture compared to standard two-dimensional surface-code architectures at $N_L = 50$ logical qubits. For fairness, we use only surface-code logical qubits in both the Pangaea architecture and the comparative two-dimensional architecture~\cite{Litinski2019}. 

\begin{figure}[!h]
    \centering
    \includegraphics[width=\columnwidth]{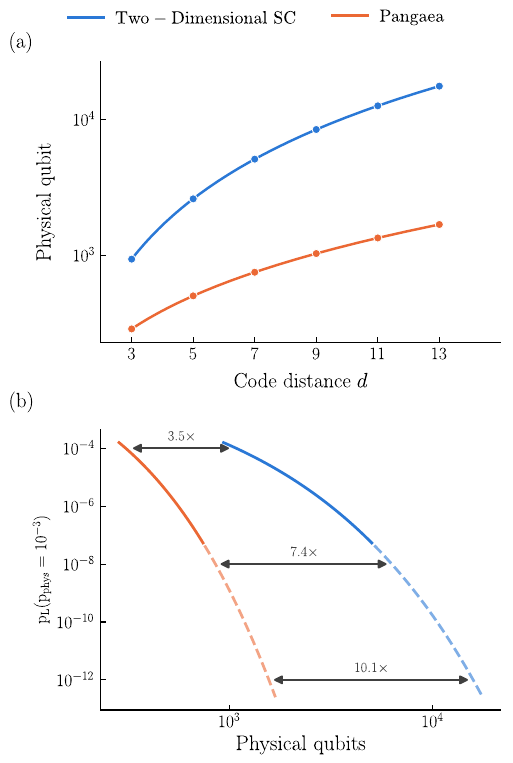}
    \caption{Physical qubit count and iso-resource comparison between two-dimensional and Pangaea architectures for the merge operation at $N_L=50$ logical qubits. \textbf{(a)} Total physical qubits for a single merge vs.\ code distance $d$: traditional ancilla regions scale as $O(d^2N_L)$, while Pangaea's bus scales as $O(dN_L)$. \textbf{(b)} Estimated logical error rate $p_\mathrm{L}$ vs.\ physical qubit count at fixed $p_{\rm phys}=10^{-3}$, obtained by re-parameterizing each curve in (a) through the fitted $p_\mathrm{L}(d,p)$ (solid: fit to simulated $d\le 7$ data; dashed: log-linear extrapolation to $d\ge 9$). Horizontal guides mark matched-$p_\mathrm{L}$ qubit-count ratios of $3.5\times$, $7.4\times$, and $10.1\times$ fewer qubits for Pangaea at $p_\mathrm{L} = 10^{-4}, 10^{-8}, 10^{-12}$ respectively.}
    \label{fig:qubit_budget_analysis}
\end{figure}

Performing long-range lattice surgery in two-dimensional architectures requires a dedicated ancilla region of $O(d^2N_L)$ qubits alongside the two $O(d^2)$ logical qubit patches. This is in contrast to the Pangaea bus mediated protocol, whose routing region grows only as $O(dN_L)$ (Appendix~\ref{app:physical_qubit_count}). 
Fig.~\ref{fig:qubit_budget_analysis}(a) shows the resulting total qubit count for both architectures. By definition, there is an $O(d)$ scaling advantage of the quantum bus in the ancilla region. It follows directly that at code distances ($d \in \{9, 11, 13\}$) this becomes an order-of-magnitude improvement in physical qubit count for identical $N_L$.

Fig.~\ref{fig:qubit_budget_analysis}(b) re-expresses the same data as an iso-resource comparison: the logical error rate of the scissors operator, $p_\mathrm{L}$, is plotted against qubit count rather than against $d$, so that the two curves can be read at matched physical qubit budget. To quantify logical error rates and how they differ between two-dimensional architectures and Pangaea, we estimate $p_\mathrm{L}(d,p_{\rm phys})$ for both. For Pangaea, we fit our simulation data at $d=3,5,7$ and extrapolate beyond it, and compare to estimates for the two-dimensional architecture. To perform a fair comparison, we assume that along the merge boundary of lattice surgery, both architectures have the same error rate. This is a fair assumption because the boundaries themselves have the same scaling behavior for both protocols, and are exactly the same at bus length $\ell=1$. However, the number of qubits used in the mediating patch is significantly different with increasing numbers of logical qubits (and therefore $\ell$). The total logical error rate is a combinatorial function of the area footprint of the ancillary regions: the larger the footprint, the more errors are generated. We model this as $p_\mathrm{L} = A(d)\,p_{\rm phys}^{(d+1)/2}$~\cite{Bodeker2026}, where the $d\ge 9$ points in Fig.~\ref{fig:qubit_budget_analysis}(b) extend the fit and should be read as order-of-magnitude estimates, not as simulated results. Thus, assuming a realistic physical error rate of $p_{\rm phys}=10^{-3}$~\cite{GoogleQuantumAI2025BelowThreshold}, and representative code distance of $d=13$, Pangaea reaches a logical error rate of $p_\mathrm{L}\approx10^{-12}$ using roughly $10\times$ fewer physical qubits than two-dimensional surface-code architectures at the same $p_\mathrm{L}$ at 50 logical qubits. 

\section{Heterogeneous Magic State Distillation}
\label{sec:15to1_distillation}

One of the main bottlenecks in fault-tolerant quantum computing is the creation of logical non-Clifford states~\cite{Bocharov2015, BravyiKitaev2005, BeverlandKubica2021, Campbell2017}, and in particular T-states~\cite{BravyiHaah2012, ChamberlandNoh2020, Hirano2025Efficient, Itogawa2024, LitinskiCostly2019, Tiurev2026, SalesRodriguez2025}. State-of-the-art approaches solve this in different ways; surface-code-based architectures can distill magic states directly using noisy injection as a starting point; however, the injected state is not prepared fault-tolerantly and may contain an undetected \(O(p)\) logical error that is not suppressed by increasing the code distance~\cite{Litinski2019, SalesRodriguez2025}. More recent approaches use magic-state cultivation, a repeat-until-success protocol that incrementally verifies and enlarges an encoded magic state using post-selected fault-tolerant checks. Cultivation can substantially reduce the qubit cost relative to conventional distillation, although its efficiency depends strongly on the protocol rejection probability~\cite{Rosenfeld2025MagicStateCultivation,Vaknin2026,Tiurev2026,Gidney2024}. In color-code-based cultivation, a subsequent grafting stage transfers the accepted state onto a larger surface-code patch for robust storage and use~\cite{Gidney2024}.

\begin{figure}[htbp]
    \centering
    \includegraphics[width=\columnwidth]{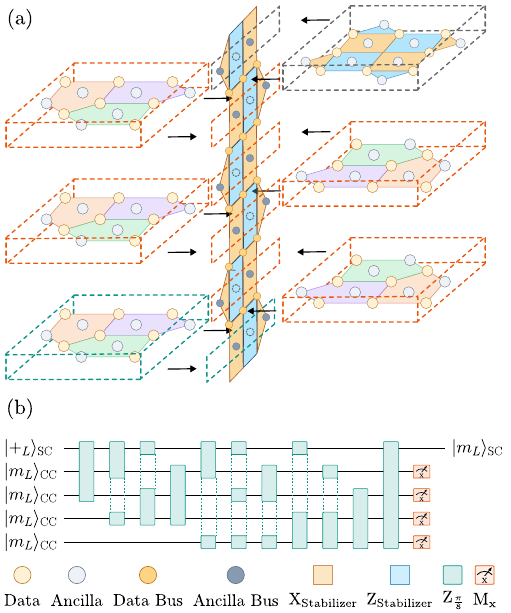}
    \caption{
    Pangaea implementation of the 15-to-1 magic-state distillation circuit. \textbf{(a)} Four color-code patches are initialized in noisy logical magic states, while the surface-code output row is initialized in \(\ket{+_L}\). The upper diagram shows the six required code patches connected via a shared quantum bus -- the surface-code output patch and five color-code patches, of which four are initialized in noisy logical magic states while the fifth serves as a color-code ancilla mediating the bus rotations. \textbf{(b)} Corresponding circuit, in which the heterogeneous \(Z\)-bus implements the eleven \(\pi/8\) Pauli-product rotations using the bus. After the eleven rotations, the four color-code patches are measured in the logical \(X\) (or, conditionally, $Y$) basis and, upon accepted outcomes, the distilled magic state remains directly encoded in the surface-code output patch.
    }
    \label{fig:Pangaea_15to1_distillation}
\end{figure}

We highlight the usefulness of the quantum bus in solving the particular problem of magic state creation using distillation. Specifically, utilizing the bus, we propose a protocol to natively distill magic states generated from heterogeneous codes: we use color-code logical qubits as noisy magic resource states, and insert a surface-code qubit into the protocol through the bus to distill the logical magic state onto. This protocol has the advantage of exploiting the strengths of color-code in generation, cultivation, and post-selection, while still resulting in a distilled surface-code logical magic state as the output. We show a schematic of our protocol in Fig.~\ref{fig:Pangaea_15to1_distillation}, along with its associated circuit. We use the standard 15-to-1 Bravyi--Kitaev \([[15,1,3]]\) Reed--Muller protocol~\cite{BravyiKitaev2005} as a benchmark, and compare to standard surface-code approaches~\cite{Litinski2019}.

We contrast the persistent protected area and handoff overhead during the distillation protocols between the two architectures. We use the standard normalization of area to one distance-\(d\) surface-code tile, and one time step to one logical lattice-surgery time step, i.e. \(d\) syndrome-extraction rounds. In this normalization, the standard surface-code implementation uses an 11-tile factory, with area \(A_{\rm SC} = 11\), for 11 logical time steps, with time \(T_{\rm SC} = 11\)~\cite{Litinski2019}. Therefore, the total space-time volume is \(V_{\rm SC} = A_{\rm SC}T_{\rm SC}=121\).

The Pangaea layout uses color-code tiles \(\ket{m_L}_\mathrm{CC}\), a single surface-code tile initialized to \(\ket{+_L}_\mathrm{SC}\), plus an additional accounting for the quantum bus.  The use of color-code resource patches also removes the dependence on additional logical ancillae required by surface-code distillation implementations ~\cite{Litinski2019}. Our construction can absorb the measurement-dependent \(P_{\pi/4}\) Clifford byproduct induced by each \(P_{\pi/8}\) rotation. The correction is absorbed directly into the destructive measurement of the reusable color-code resource patch: conditioned on the decoded bus-parity outcome, the patch is measured in either the logical \(X\) or \(Y\) basis. The logical-\(Y\) readout is implemented locally by a transversal Clifford basis change followed by destructive logical-\(X\) measurement, while the remaining Pauli byproduct is retained in the logical Pauli frame. Consequently, the adaptive correction requires neither an additional protected logical patch nor a further bus-mediated lattice-surgery step, and each of the eleven rotations occurs in one logical time step.

Our setup requires a logical-qubit area $A_\mathrm{CC} = 5$ and a surface-code area of $A_\mathrm{SC} = 1$. However, since each tile is $O(d^2)$, and our bus requires area of $O(dN_L)$, we require additional $1/d$ tiles per additional logical qubit in the bus, yielding a total area of $A_\mathrm{bus} = 5/d$ tiles, given $\ell=5$ in our protocol. We note that both surface-code and color-code tiles scale as $O(d^2)$, and in the high-distance $d$ limit dominate over the $1/d$ contribution of the bus. Therefore, the total persistent area footprint of the Pangaea distillation is $A_\mathrm{P} \approx 6$.

The time required to implement each bus-mediated \(\pi/8\) rotation is scheduled in one logical time step, and so $T_\mathrm{P}=11$. This is the same as in standard surface-code architectures~\cite{Litinski2019}. We have already shown in previous sections that bus mediation is on the same timescale as standard lattice surgery. The total resulting space-time cost is therefore $V_\mathrm{P} = A_\mathrm{P}T_\mathrm{P} = 66$, roughly half that of the two-dimensional surface-code architecture.

This highlights the basic architectural advantage of the heterogeneous implementation: the four initial magic-state logical qubits can be supplied by color-code patches optimized for injection. The output is produced directly in a surface-code patch, and the eleven remaining rotations use color-code resource states. Together, these yield a distillation protocol that significantly reduces the resources required to implement it while simultaneously eliminating the need for a post-distillation state transfer or code switch.

\section{Outlook}
\label{sec:functional_modules}

Large-scale fault-tolerant quantum computing requires three main capabilities beyond error correction to be practically useful: (i) non-Clifford gates, (ii) fault-tolerant two-qubit gates, and (iii) scalable routing between logical qubits. The Pangaea architecture answers these criteria through its implementation of the quantum bus. By decomposing long-range interactions into local measurements via gauge generators, we extend lattice surgery to a fundamental, three-dimensional fault-tolerant primitive. We have shown how to scalably connect many logical qubits to each other, enabling fault-tolerant measurements along the bus with significantly fewer qubits than two-dimensional constructions. The quantum bus's native ability to connect different code families allows the assignment of separate resources for separate tasks to which they are best suited---the basis of scalable computing principles. These features establish the bus as a composable, long-range, code-agnostic interconnect, exemplified by our heterogeneous magic state distillation protocol.

This architectural modularity is a consequence of the quantum bus verified in this work, and it comes with open problems beyond the presented results. The fault-tolerance and threshold calculations shown here are sufficient conditions for the specific surface--surface and surface--color layouts; extending them to high-rate planar tile and qLDPC codes~\cite{Steffan2025TileCodes,Gu2026NearestNeighbour,Liang2025PlanarQLDPC}, long-range multi-bus networks~\cite{Choe2024}, or dedicated subroutine blocks such as the quantum Fourier transform~\cite{Aumann2026QFT}, requires significant additional work. Furthermore, machine learning-based dedicated decoders for heterogeneous codes and multi-module constructions could be particularly useful and applicable to our architecture~\cite{tong2026learning}. 

Although the quantum bus requires local, degree-6 connectivity, realizing repeated vertical bus slots will require overcoming significant engineering challenges in practice. Its three-dimensional connectivity demands high-yield inter-layer couplers with sufficient isolation from noise. The stacked geometry must preserve access for qubit control without creating fabrication bottlenecks between logical patches and the bus. Candidate implementations include multilayer wiring and flip-chip integration~\cite{Rosenberg20173DIntegration}, through-silicon vias (TSVs)~\cite{Wang2011TSVQubits}, and modular chip-to-chip couplers~\cite{Kurpiers2018QuantumBus, Axline2018Interconnect}. However, fabrication variability, correlated noise~\cite{Gambetta2017Building, McEwen2021ResonatorLoss} and their impact on performance must be characterized experimentally. Establishing scalable fabrication procedures for these repeated three-dimensional interfaces is therefore an important prerequisite for translating the advantages derived here into a large-scale superconducting processor.

Nonetheless, this work provides a blueprint for large-scale fault-tolerant quantum computing. Pangaea makes the quantum bus the main building block of its architecture, where a network of such buses serves as a fault-tolerant interconnect across modular components. Our work complements ongoing research in quantum architecture design across competing modalities~\cite{Bluvstein2024, Yoder2025, BravyiCross2024, RyanAnderson2024Teleportation,Cain2026Shor}, underscoring how such improvements are becoming necessary to alleviate core bottlenecks in quantum algorithms~\cite{Shor1994, Coppersmith2002, Gidney2025FactorRSA}. Future work will extend these principles to higher-rate codes and more efficient decoders, alongside demonstrations of quantum bus implementations in superconducting quantum hardware.

\section*{Acknowledgments}
We would like to thank Mor M. Roses, Guy Koren, Daniel Dahan, Amir Dalal, Naftali Kirsh and Samuel D. Escribano for feedback on initial versions of this manuscript. NK acknowledges support of ISF-Quantum and EU OpenSuperQPlus projects. 

\bibliography{bibliography}

\appendix

\section{Surface-Code and Color-Code Definitions}
\label{app:code_definitions}

\subsection{The Surface-Code}
\label{subsec:SC_app}
The surface-code is a two-dimensional topological stabilizer code defined on a local lattice of physical qubits~\cite{Dennis2002,Horsman2012,Litinski2019, GoogleQuantumAI2025BelowThreshold}. In the \(n,k,d \) notation, we can describe this code as a \([[n=d^2,k=1,d]]\) code, where $n$ is the number of qubits, $k$ is the number of logical qubits, and $d$ is the distance of the code~\cite{Gottesman1997Stabilizer}. In a planar implementation, data qubits are placed on the edges or vertices of a two-dimensional lattice, and the code space is the simultaneous \(+1\) eigenspace of local commuting stabilizers associated with vertices and faces. We write these as
\begin{equation}
    S^{(X)}_v = \prod_{i\in v} X_i,
    \qquad
    S^{(Z)}_p = \prod_{i\in p} Z_i,
    \label{eq:surface_stabilizers}
\end{equation}
where \(S^{(X)}_v\) is a vertex, or star, operator acting on qubits adjacent to the vertex \(v\), and \(S^{(Z)}_p\) is a plaquette operator acting on qubits around the face \(p\). Boundary checks may have reduced weight, but obey the same Pauli type convention.

Logical operators are represented by non-contractible Pauli strings connecting opposite boundaries of the planar patch. For a single encoded qubit we use representatives
\begin{equation}
    \overline{X}=\prod_{i\in\partial_X} X_i,
    \qquad
    \overline{Z}=\prod_{i\in\partial_Z} Z_i,
    \label{eq:surface_logicals}
\end{equation}
where \(\partial_X\) and \(\partial_Z\) are paths connecting the corresponding rough/smooth boundary pairs. Multiplication by stabilizers deforms these paths without changing their logical action. The surface-code distance \(d\) is the minimum weight of any nontrivial representative of \(\overline{X}\) or \(\overline{Z}\).

\subsection{The Color-Code}
\label{subsec:CC_app}

The color-code is a two-dimensional topological stabilizer code defined on a trivalent lattice with three-colorable faces~\cite{Bombin2006,Lacroix2025}. In the \(n,k,d \) notation, we can describe this code as a \([[n=(3d^2+1)/4,k=1,d]]\) code.
Physical qubits are placed on lattice vertices, and each face \(f\) supports both
an \(X\)-type and a \(Z\)-type stabilizer,
\begin{equation}
    S^{(X)}_f = \prod_{i\in f} X_i,
    \qquad
    S^{(Z)}_f = \prod_{i\in f} Z_i,
    \label{eq:color_stabilizers}
\end{equation}
where the products run over the qubits incident on \(f\). The simultaneous \(+1\) eigenspace of all face checks defines the encoded subspace. Because the
same face supports both Pauli types, the color-code stabilizer structure is symmetric under exchange of \(X\) and \(Z\).

For a planar color-code patch encoding one logical qubit, logical Pauli operators
are string operators connecting appropriate colored boundaries,
\begin{equation}
    \overline{X}=\prod_{i\in\partial} X_i,
    \qquad
    \overline{Z}=\prod_{i\in\partial'} Z_i,
    \label{eq:color_logicals}
\end{equation}
where \(\partial\) and \(\partial'\) are nontrivial paths chosen so that the two operators anticommute. The color-code distance \(d\) is the minimum weight of any nontrivial logical representative. The symmetry between \(X\)- and \(Z\)-type checks permits transversal implementations of some Clifford operations, including the logical Hadamard, which can be used to exchange the exposed boundary basis. 

\section{Physical Qubit Count}
\label{app:physical_qubit_count}

In our construction of the bus, the bus has a width $d$ to match the distance of the logical qubits connected to it. Furthermore, we define the bus length $\ell$ as the number of gauge data-qubit rows in the bus; a length-$\ell$ bus connects $N_L=\ell+1$ logical qubits. Note that, for the smallest such value $\ell=1$, we recover standard lattice surgery connecting two logical qubits. In our construction, we consider buses of odd length $\ell$, such that we can define each bus as uniquely connecting a single type of logical Pauli operator for all logical qubits on the bus.

We start by calculating the number of data qubits required to construct the bus. Notice that, by definition, we need $\ell$ layers of data qubits for a length $\ell$ bus. Furthermore, we need $d$ data qubits in each data row to match the distance of the logical qubits. Therefore, we need a total of $n_\mathrm{data} = d \ell$ data qubits in the bus. We now calculate the number of supporting ancillas required to operate the bus. By construction, the bus has missing qubits as ``slots'' into which the rough boundaries of the logical qubits are inserted. The ancilla qubits form rows interleaved with the data rows, so a length-$\ell$ bus contains $\ell+1$ ancilla rows---one more than the number of data rows, with the extra row arising from the rough boundary of the bus itself---and each ancilla row contains $(d+1)/2$ ancillas. Therefore we have a total of $n_\mathrm{ancilla} = \left(\ell+1\right) \left(d+1\right)/2$. This can be seen explicitly in Fig.~\ref{fig:bus_growth}(a). The total number of qubits in the bus is
\begin{equation}
\begin{aligned}
    n_\mathrm{bus} 
    &= n_\mathrm{ancilla} + n_\mathrm{data} \\
    &= \left(\ell+1\right) \left(\frac{d+1}{2}\right) + d\ell = \frac{3d\ell+d+\ell+1}{2}.
\end{aligned}
\end{equation}

We find that the number of physical qubits required to implement our quantum bus scales as $O(dN_L)$, whereas standard two-dimensional ancilla patch constructions require $O(d^2N_L)$~\cite{Litinski2019}. 

\section{Detector Model for Threshold calculations}
\label{app:detector_model}
Detector events are defined as changes between consecutive measurements of an active stabilizer or gauge check, with bridge detectors inserted when checks are temporarily disabled during the merge and then restored, comparing the last pre-merge outcome to the first post-merge outcome. For this comparison to be well defined, the \(X\)-type checks that border the merge region are measured once more immediately before the bus qubits are reset, so that each of them has a recorded pre-merge outcome to bridge against; omitting this measurement would leave those checks with no last pre-merge outcome at all once the bus is reset, an undetected fault path of weight less than \(d\) that would otherwise break the circuit-level distance of the merge. The logical observable sampled by the detector sampler is the parity of the first-merge bus \(X\)-check outcomes, multiplied by the final \(X\)-basis readout on the qubits supporting the \(X_\mathrm{L}^{(L_0)}X_\mathrm{L}^{(L_1)}\) representative. A logical failure is counted when the decoder's predicted observable differs from this sampled observable.

Each protocol is swept over distances \(d\in\{3,5,7\}\), bus lengths \(\ell\in\{1,3,5,7\}\), and physical error rates \(p_{\rm phys}\in \{3\times10^{-4},\,10^{-3},\,3\times10^{-3},\,10^{-2},\,3\times10^{-2},\, 10^{-1},\,3\times10^{-1}\}\), each using \(r_{\rm pre\text{-}op}=r_{\rm merge}=r_{\rm post\text{-}op}=d\) syndrome-extraction rounds. Surface-to-surface circuits are decoded with minimum-weight perfect matching (MWPM) on the fully decomposed detector error model. Surface-to-color circuits are decoded with belief-propagation plus ordered-statistics decoding (BPOSD) on the un-decomposed model, which retains the hyperedge faults generated by \(Y\)-type data-qubit errors that flip more than two detectors. The BPOSD stage uses product-sum message updates, a parallel (synchronous) schedule, a maximum of 20 BP iterations, and OSD order 10 with the combination-sweep method. These were chosen empirically over the weaker settings of order 0 with a serial schedule, since low-order OSD post-processing is a documented source of non-monotonic logical error rates as \(p_{\rm phys}\) varies, and the serial schedule produced a non-monotonic collapse in \(p_\mathrm{L}\) at \(d=5\) for select values of \(p_{\rm phys}\). Raising the OSD order and switching to the parallel schedule removed both artifacts, with the parallel schedule also converging substantially faster at that distance.

The number of shots sampled per point is adaptive rather than fixed, so that a genuinely small logical error rate is not mistaken for insufficient sampling. A floor number of shots is always drawn first, ranging from \(10^5\) shots at the largest \(p_{\rm phys}=3\times10^{-1}\) up to \(2\times10^7\) shots (MWPM, \(d=7\)) or \(6\times10^6\) shots (BPOSD, \(d=7\)) at the smallest \(p_{\rm phys}=3\times10^{-4}\), scaling up with \(d\) to compensate for the smaller logical error rate at higher distance; this floor is further multiplied by \(1+0.15(\ell-1)\), capped at \(2\times\), for longer bus lengths. Beyond the floor, shots continue to be drawn in batches until at least 20 logical failures are observed for the surface-to-surface curves, or at least 50 for the surface-to-color curves, up to a hard ceiling of \(2000\times\) the floor (capped at \(10^{11}\) total shots for MWPM, \(10^{9}\) for BPOSD) or a wall-clock limit of 30 minutes per point. A point that reaches either limit before its target failure count is retained with the appropriate statistics, rather than treated as converged.

For each simulated point we estimate \(p_\mathrm{L}=\Pr[\hat{o}\ne o]\), where \(o\) is the sampled logical observable and \(\hat{o}\) is the decoder's prediction, with confidence interval \(p_\mathrm{L}\pm1.96\sqrt{p_\mathrm{L} (1-p_\mathrm{L})/n_{\rm shots}}\), clipped to \([0,1]\). The pseudo-threshold \(p^*\) reported in Fig.~\ref{fig:thresholds} is a finite-distance crossing estimate, obtained by linearly interpolating between the two sampled physical error rates at which the sign of \(p_\mathrm{L}-p_{\rm phys}\) changes, rather
than by fitting an asymptotic threshold model.

\section{Native Multi-Qubit Bus Measurements}
\label{app:multi_terminal_bus}
\begin{figure}[htbp]
    \centering
    \includegraphics[width=0.7\columnwidth]{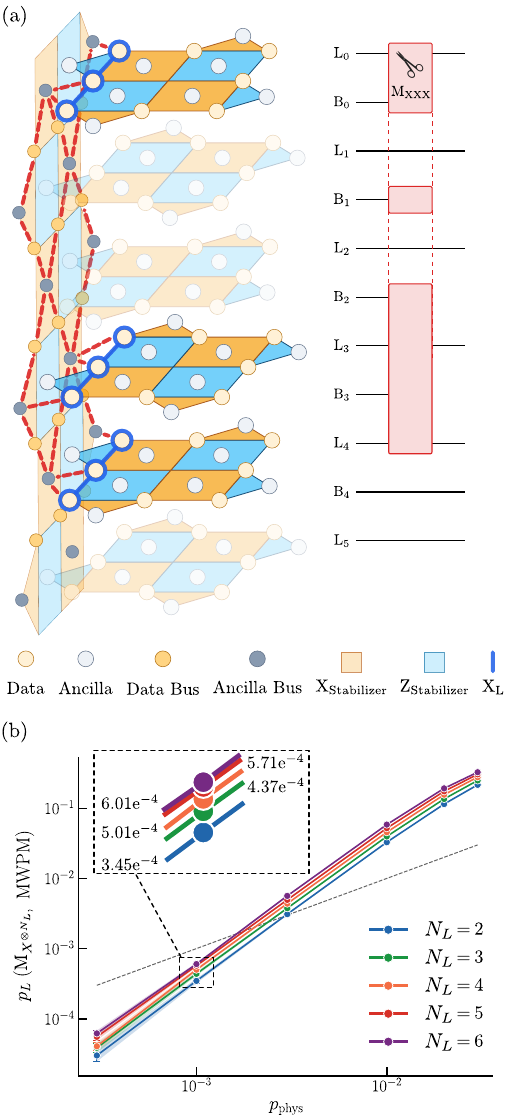}
    \caption{An illustration of a native three-qubit $M_{XXX}$ operation through the quantum bus. \textbf{(a)} A quantum bus of length 5 ($\ell$) connecting 6 ($N_L$) logical surface-code qubits. The three highlighted distance-3 patches each represent a logical qubit participating in the interaction; dashed red lines mark the qubits participating in the gauge checks mediating the measurement, and blue lines highlight the three logical Pauli-$X$ operators involved. \textbf{(b)} Simulated logical error rate $p_\mathrm{L}$ vs. physical error rate $p_{\rm phys}$ for a varying number of jointly measured bus terminals $(N_L=2$--$6$, $d=3$, bus length 9, MWPM decoding). $p_\mathrm{L}$ increases monotonically with $N_L$, as expected, but stays within the same order of magnitude across the full range.}
    \label{fig:multi_qubit_gates}
\end{figure}

Our quantum bus can natively perform multi-qubit Pauli measurements: a single measurement of an $N_L$-body Pauli product, rather than a sequence of pairwise lattice-surgery merges to decompose the operator. Fig.~\ref{fig:multi_qubit_gates}(a) illustrates the protocol for a native $M_{XXX}$ measurement using an example length-5 bus of distance-3 surface-code qubits. To showcase fault-tolerance of our protocol, we compute the logical error rates of $N_L$-body Pauli measurement with $N_L\in \{2, \ldots, 6\}$  distance-3 surface-code qubits, with constant bus length $\ell=9$. Fig.~\ref{fig:multi_qubit_gates}(b) reports the logical error rate using MWPM decoding. As the number of jointly measured qubits ($N_L$) grows from 2 to 6, we see that $p_\mathrm{L}$ increases monotonically with $N_L$, as expected, since additional logical qubits provide more places for a fault to enter. However, logical error rates stay within the same order of magnitude across the full range.

\begin{figure}[htbp]
    \centering
\includegraphics[width=0.7\columnwidth]{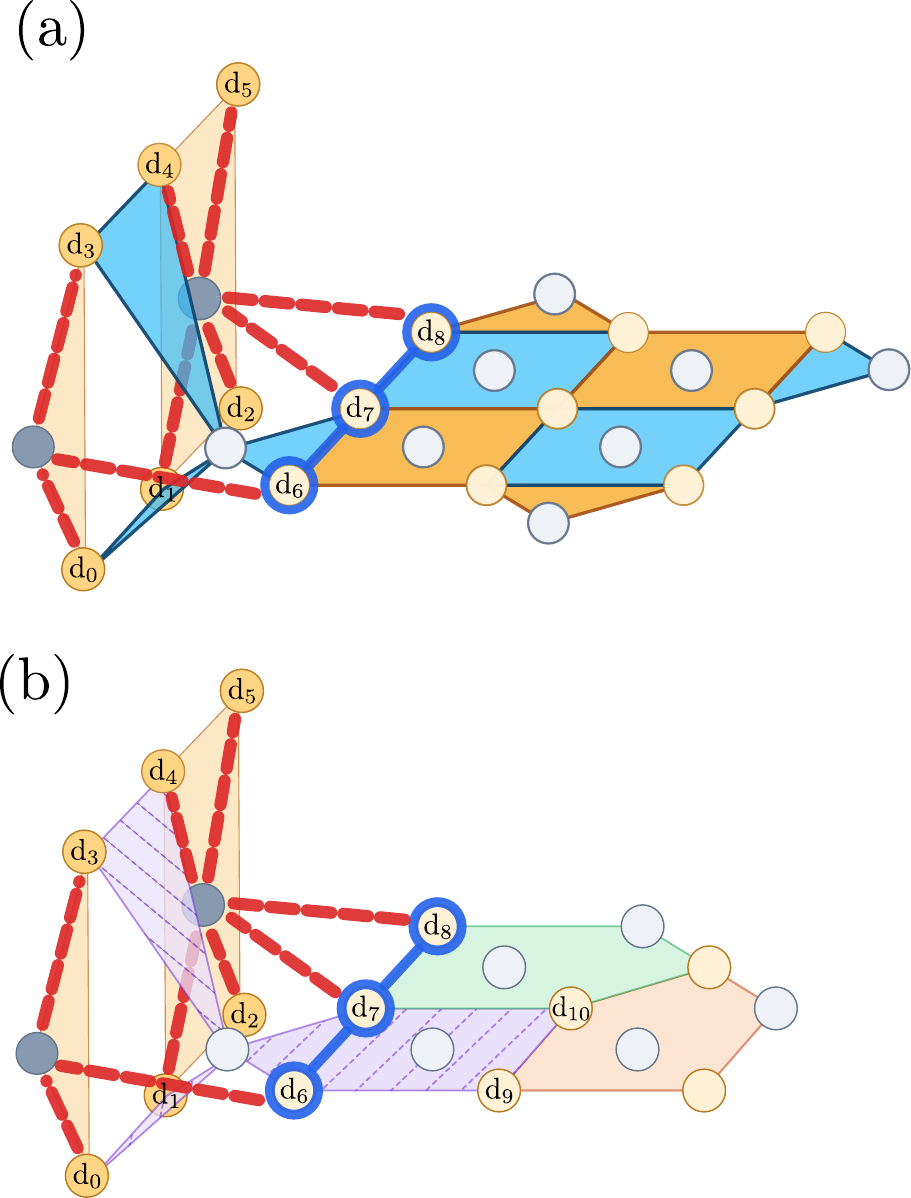}
    \caption{
    A middle logical-qubit of distance $d=3$ inserted into a single bus slot: \textbf{(a)} surface-code logical qubit \textbf{(b)} color-code logical qubit
    }
    \label{fig:multi_Z_fix}
\end{figure}

A problem arises when a patch is inserted in the middle of the bus rather than at an endpoint. This can be seen in the weight-6 $X$-stabilizers shown in Fig.~\ref{fig:multi_Z_fix}(a) for a surface-code patch, and Fig.~\ref{fig:multi_Z_fix}(b) for a color-code patch. This stabilizer uses qubits both above and below it in the bus. In the surface-code case, the corresponding weight-6 $Z$-stabilizers of the same bus slot no longer commute, as they overlap on exactly three data qubits. In general, the same issue would arise in a $Z$-bus with complementary weight-6 $X$-stabilizers. We show an example explicitly in Fig.~\ref{fig:multi_Z_fix}(a) on the surface-code patch with logical operator $\overline{X_L} = X_6X_7X_8$. One can immediately see that the weight-6 $\mathcal{G}^{(X)}_{B,0}=X_1X_2X_4X_5X_7X_8$ and weight-3 $\mathcal{G}^{(X)}_{B,1}=X_0X_3X_6$ stabilizers do not commute with $\mathcal{G}^{(Z)}_{B,0}=Z_0Z_1Z_3Z_4Z_6Z_7$. The reason is that $\mathcal{G}^{(Z)}_{B,0}$ overlaps with both stabilizers an odd number of times: on $d_1, d_4, d_7$ ($d_0, d_3, d_6$) for $\mathcal{G}^{(X)}_{B,0}$ ($\mathcal{G}^{(X)}_{B,1}$) respectively.  We address this by redefining the $Z$ stabilizers for this patch: we break apart the weight-6 stabilizer into two separate weight-4 stabilizers, which reuse exactly 2 qubits, therefore canceling their contribution. We define these weight-4 stabilizers so that they both commute with the weight-6 and weight-3 $X$-stabilizers (now intersecting on two qubits instead of three). 
Using the example in Fig.~\ref{fig:multi_Z_fix}(a), this means we measure two weight-4 stabilizers $\hat{\mathcal{G}}^{(Z)}_{B,0}=Z_0Z_1Z_6Z_7$ and $\hat{\mathcal{G}}^{(Z)}_{B,1}=Z_3Z_4Z_6Z_7$. Thus, all weight-6 stabilizers intersect on an even number of data qubits, which corrects the original issue, resulting in the stabilizer group of the merged code \(\mathcal{G}_{B,k}^{(Z)}\) commuting with \(\mathcal{G}_{B,k}^{(X)}\). Note that this correction arises only when performing multi-qubit Pauli measurements, as in the two-qubit case the problematic weight-6 stabilizers always reduce to weight-4, and commute appropriately. 

For the color-code patches in Fig.~\ref{fig:multi_Z_fix}(b), with the same logical operator $\overline{X_L} = X_6X_7X_8$, the weight-4 Z purple color-code stabilizer ($Z_6Z_7Z_9Z_{10}$) naturally expands to a weight-8 Z stabilizer with the four bus qubits $\mathcal{G}^{(Z)}_{B,0}=Z_0Z_1Z_3Z_4Z_6Z_7Z_9Z_{10}$, and also does not commute with the $X$-stabilizers. In this case we take a similar approach and split the non-commuting weight-8 $Z$-stabilizer into two weight-6 $Z$-stabilizers - \(\hat{\mathcal{G}}_{B,0}^{(Z)}=Z_0Z_1Z_6Z_7Z_9Z_{10}\) and \(\hat{\mathcal{G}}_{B,1}^{(Z)}=Z_3Z_4Z_6Z_7Z_9Z_{10}\) which now commute with the merged weight-6 and weight-3 X stabilizers as they now overlap on an even number of data qubits (\(\hat{\mathcal{G}}_{B,0}^{(Z)}\) ($\hat{\mathcal{G}}_{B,1}^{(Z)}$) and $\mathcal{G}^{(X)}_{B,0}$ ($\mathcal{G}^{(X)}_{B,1}$) overlap on $d_4, d_7$ ($d_3, d_6$)).

\section{Gauge-Fixing Degrees of Freedom in the Bus}
\label{app:gauge-fixing-bus}

Here we verify that the quantum bus does not introduce any unconstrained logical degrees of freedom. We do this by counting the number of degrees of freedom for the three possible code combinations in the bus: the two homogeneous cases, all-surface and all-color, and the heterogeneous surface--color case. In subsystem lattice surgery, it is known that color-code qubits lose a stabilizer check at the boundary with the merge region~\cite{Nautrup2017}, resulting in an additional degree of freedom. Surface-code qubits' stabilizers extend naturally into the bus as in standard lattice surgery, and therefore do not lose any checks. We therefore examine the color-code as the basis of our analysis. In general, each distance-$d$ color-code logical qubit, whether at the end of the bus or in its interior, loses $(d-1)/2$ independent local boundary checks during the merge. These checks are absorbed into extended seam generators spanning the bus in order to ensure commutation of all stabilizers in the merged bus region. Each dropped check introduces an additional degree of freedom, corresponding to one gauge qubit~\cite{Vuillot2019GaugeFixing}. For $d=3$, each color-code patch loses exactly one such local boundary check.

Consider $N_L$ logical qubits $L_0,\ldots,L_{N_L-1}$ joined by a length-$\ell$ bus of width $d$, of which $\mu$ are color-code patches and $N_L-\mu$ are surface-code patches. When $\mu=0$ or $\mu=N_L$, the construction reduces to the homogeneous all-surface or all-color case, respectively, while $0<\mu<N_L$ describes the heterogeneous case. We use the $[[n,k,r,d]]$ notation for a subsystem code of $n$ data qubits, $k$ protected logical qubits, $r$ gauge qubits, and distance $d$, with stabilizer-center rank
\begin{equation}    
    s=n-k-r.
\end{equation}

Before the merge (pre-merge) and after the destructive bus readout (post-merge), the bus qubits are independent. The split configuration therefore contains only the data qubits $(n_{\mathrm{split}})$ belonging to the logical patches,
\begin{equation}    
    n_{\mathrm{split}}
    =
    \sum_{i=0}^{N_L-1} n_i,
\end{equation}

where
\begin{equation}
    n_{\mathrm{SC}}(d)=d^2,
    \qquad
    n_{\mathrm{CC}}(d)=\frac{3d^2+1}{4}
\end{equation}
for the surface-code and color-code described in App.~\ref{app:code_definitions}. Each patch supplies its complete stabilizer group and encodes one logical qubit, so that
\begin{equation}
\begin{aligned}
    s_{\mathrm{split}}
    &=
    \sum_{i=0}^{N_L-1}
    \operatorname{rank}\mathcal S_i \\
    &=
    \sum_{i=0}^{N_L-1}(n_i-1) \\
    &=
    n_{\mathrm{split}}-N_L.
\end{aligned}
\end{equation}
Since the split configuration drops no checks, it contains no gauge qubits, and therefore $r_{\mathrm{split}}=0$. The number of logical qubits is
\begin{equation}    
    k_{\mathrm{split}}
    =
    n_{\mathrm{split}}
    -s_{\mathrm{split}}
    -r_{\mathrm{split}}
    =
    N_L.
\end{equation}
The separated system is therefore described by the subsystem code
\begin{equation}
    \left[\left[
    n_{\mathrm{split}},\,
    N_L,\,
    0,\,
    d
    \right]\right],
\end{equation}
which recovers $[[n_{\rm split}, N_L, d]]$ in the standard $[[n, k, d]]$ notation. During the merge, the $d\ell$ bus data qubits become part of the active code, and hence the total number of data qubits is
\begin{equation}    
\begin{aligned}
    n_{\mathrm{merge}}
    &= n_{\rm split} + d\ell = 
    \sum_{a=0}^{N_L-1}n_a+d\ell \\
    &=
    (N_L-\mu)d^2
    +\mu\frac{3d^2+1}{4}
    +d\ell.
\end{aligned}
\end{equation}

During the merge, the $\mu$ color-code logical qubits contribute
\begin{equation}
    r_{\mathrm{merge}}
    =
    \mu\,\frac{d-1}{2}
\end{equation}
gauge qubits in total, independent of where the color-code logical qubits are positioned along the bus. The activated bus checks measure one independent $N_L$-body dressed logical operator,
\begin{equation}
    M_{P^{\otimes N_L}}
    =
    \prod_{a=0}^{N_L-1}\overline P^{(a)}.
\end{equation}
This measurement imposes one independent logical constraint and therefore reduces the number of protected logical degrees of freedom from $N_L$ to $k_{\mathrm{merge}}=N_L-1$.

The rank of the merged code is consequently
\begin{equation}
\label{eq:merge_gauge_form_first}
\begin{aligned}
    s_{\mathrm{merge}}
    &=
    n_{\mathrm{merge}}
    -k_{\mathrm{merge}}
    -r_{\mathrm{merge}} \\
    &=
    n_{\mathrm{merge}}
    -N_L+1
    -\mu\frac{d-1}{2}.
\end{aligned}
\end{equation}

Equivalently, in terms of the separated stabilizer rank,
\begin{equation}
    s_{\mathrm{merge}}
    =
    s_{\mathrm{split}}
    +d\ell
    +1
    -\mu\frac{d-1}{2}.
\end{equation}
\vspace{0.5\baselineskip}

The term $d\ell$ accounts for the bus qubits introduced into the active merged code, the additional $+1$ represents the measured joint logical constraint, and the final term accounts for the stabilizers converted into gauge degrees of freedom at the color-code interfaces. The merged configuration is therefore the subsystem
code~\cite{Nautrup2017,ThomsenKesselring2022}

\begin{equation}
\label{eq:merge_gauge_form}
    \left[\left[
    n_{\mathrm{merge}},\,
    N_L-1,\,
    \mu\frac{d-1}{2},\,
    d
    \right]\right].
\end{equation}
\vspace{0.5\baselineskip}

For $\mu=0$, the construction recovers the ordinary all-surface stabilizer code

\begin{equation}
    \left[\left[
    n_{\mathrm{merge}},\,
    N_L-1,\,
    d
    \right]\right],
\end{equation}

and for $\mu=N_L$, we obtain
\begin{equation}
    \left[\left[
    n_{\mathrm{merge}},\,
    N_L-1,\,
    N_L\frac{d-1}{2},\,
    d
    \right]\right].
\end{equation}

The measurements of the activated bus gauge checks fix the code to the intended joint observable up to the bus-local operator $S_B^{(P)}$. The final destructive bus readout at the split fixes $S_B^{(P)}$ together with the $r$ gauge qubits, restoring every logical qubit's original stabilizer group -- including, at each color-code, the one local check that was dropped during the merge -- and returning the system to $N_L$ separate distance-$d$ patches with the measured joint logical parity retained as a classical outcome.

\end{document}